# A scalable and resource-efficient pipelined p-computer for probabilistic Ising machines

*Deborah Volpe[1], Eleonora Raimondo[1], Andrea Grimaldi[2], Pedram Khalili Amiri[3,4], Stefano Chiappini[1], Anna Giordano[5], Mario Carpentieri[2], Hyunsoo Yang[6], Massimo Chiappini[1] and Giovanni Finocchio[1,7]*∗*

[1] Istituto Nazionale di Geofisica e Vulcanologia, Rome, 00143, Italy.

[2] Department of Electrical and Information Engineering, Politecnico di Bari, Bari, 70126, Italy.

[3] Department of Electrical and Computer Engineering, Northwestern University, Evanston, IL 60208, United States of America.

[4] Applied Physics Program, Northwestern University, Evanston, IL 60208, United States of America.

[5] Dipartimento di Ingegneria, Università degli Studi di Messina, Messina, 98166, Italy

[6] Department of Electrical and Computer Engineering, National University of Singapore, Singapore

[7] Dipartimento di Scienze Matematiche e Informatiche, Scienze Fisiche e Scienze della Terra, Università degli Studi di Messina, Messina, 98166, Italy

∗ Corresponding author: gfinocchio@unime.it

Funding: D.V. and E.R. also acknowledge funding from the project PON Capitale Umano (CIR_00030) funded by the Italian Ministry of Universities and Research (MUR), and D.D. n. 47 del 20 febbraio 2025 – “Assunzione di ricercatori internazionali post-dottorato”, CUP: D53C25000630007 (D.V.) and CUP: D53C25000620007 (E.R.).

## Abstract

Probabilistic Ising machines (PIMs) based on probabilistic bits offer a hardware-friendly route to solve combinatorial optimization problems, but most digital implementations achieve high throughput by exploiting sparse interactions. This limits their applicability to dense problems, for which memory bandwidth and data movement become the dominant bottlenecks. Here, we show a resource-efficient pipelined Field-Programmable Gate Array architecture enabling high-throughput execution of fully-connected PIMs while maintaining scalability and modularity. This architecture design combines a deeply pipelined (>20 stages) probabilistic bit update path, which overlaps spin evaluation and local-field updates, with a bandwidth-aware on-chip memory organization for the coupling and bias matrices. The architecture supports 512 p-bits with 16-bit fixed-point coefficients and 1024 and 2048 p-bits with 10-bit and 2-bit coefficients, respectively, and operates at up to 300 MHz. At fixed degree of parallelization, it delivers an order-of-magnitude higher update rate than an optimized non-pipelined baseline, while improving the time-area trade-off for dense workloads. Validation on portfolio optimization and low-density parity-check decoding shows close agreement with software references and substantial reductions in time-to-solution relative to the non-pipelined design, establishing pipelining as an effective route to scalable digital probabilistic computing for dense optimization problems.

## Introduction

Many industrially relevant problems in science and engineering[1], spanning communication networks[2], bioinformatics[3], logistics[4], and artificial intelligence[5], rely on the solution of combinatorial optimization problems (COPs). Many of these problems are nondeterministic polynomial-time (NP)-hard, motivating the development of alternative computing paradigms capable of exploring large and complex energy landscapes[6–9] with entropic barriers[10]. Among approximate solvers for COPs, Ising machines (IMs) have emerged as a promising computational paradigm over the past decade, with implementations based on quantum annealers[11,12], oscillator networks[13–15], simulated bifurcation machines[16–18], and probabilistic architectures[19,20]. A fundamental challenge in advancing IMs is the design of hardware architectures that are both scalable and resource-efficient in terms of area occupancy and energy consumption, while preserving or improving performance for the search of high-quality solutions[21]. Probabilistic Ising machines (PIMs) based on probabilistic bits (p-bits) have emerged as a promising paradigm, its flexible and hardware-friendly formulation allows their deployment in fully digital[22], analog[23], and hybrid analog-digital architectures[24]. PIMs emulate the stochastic sampling of spin systems by using pseudo[25] or true random number generation[23], allowing them to approximate Boltzmann distributions. State-of-the-art Complementary Metal-Oxide-Semiconductor (CMOS) and Field-Programmable Gate Array (FPGA) PIM implementations[18–21] prove a degree of flexibility through annealing algorithms[22], high throughput and competitive performance across sever optimization tasks[26–29].

Most existing PIM accelerators exploit sparsity in the coupling matrix to reduce memory and bandwidth requirements. However, many industrially relevant Ising formulations, such as Low-Density Parity-Check (LDPC) and portfolio optimization problems, are inherently dense[30–32]. Consequently, architectures that rely on fixed topological connectivity (e.g.

Chimera, Pegasus, and Zephyrus[33]) or sparsification[25] lose much of their advantage on these dense problems. Both approaches require additional Ising spins for graph embedding that may modify the original problem energy landscape by adding metastable states[33,34], as well as costly problem-specific pre-processing and heuristics that add computational overhead. In addition, depending on the number of spin parallel updates planned additional hardware resources are necessary. Such limitations motivate the design of a scalable hardware supporting fully-connected problems natively.

Here, we design, synthetize, and experimentally validate a scalable, low latency, and high throughput PIM architecture with deeply configurable pipelined update engine (more than 20 stages) with hierarchical organization that supports all-to-all interactions and sustains high throughput without exploiting sparsity. Our design guarantees the following characteristics:

(i) High operating frequency up to 300 MHz (AMD ZCU106 board) that is one order-of-magnitude larger than an optimized non-pipelined baseline (40 MHz) with comparable logic and memory utilization and parameters configuration;

(ii) Efficient memory organization with bandwidth-aware access schemes that combines shared and paged memories with multi-row banking and symmetry-aware reuse;

(iii) Logic circuit design offering a parameterizable numerical precision (configurable bit-width for both integer and fractional components);

(iv) Adjustable Degree Of Parallelization (DOP) defined as the number of $J$-coefficients that can be read and processed per clock cycle, (i.e. the number of local-field updates that can start concurrently);

(v) A configurable pipeline depth where arithmetic operations are deeply pipelined and local-field updates are partially overlapped across spins with logical ordering of state updates strictly preserved.

The quality of the hardware implementation is compared with a software PIM[22] on invertible logic and Max-Cut instances. In addition, we benchmarked and evaluated our pipelined architecture with 23 pipeline stages with (i) portfolio optimization[31,35] and (ii) LDPC[30] decoding, the former relevant in finance, the latter used for error correction in telecommunication and hard-disk drives. For both problems, our results show a speedup of up to one order of magnitude over the non-pipelined optimized implementation for an equal DOP. This improvement results from a combination of faster clock frequency and an optimized update schedule of the local fields, which enables an overlap without stalls between local-field updates and state-evaluation phases. This strategy enables a time reduction beyond what is achievable through clock-frequency scaling alone. In addition, this design reduces FPGA logic utilization by up to a factor of 16 for a fixed size of the problem. These results enable this pipelined architecture as a promising template for scalable PIM accelerators and as a viable design strategy toward future digital Application-Specific Integrated Circuit (ASIC) and hybrid implementations combining CMOS with optics or spintronics targeting dense, real-world optimization workloads.

**Background on probabilistic Ising machines and validation problems**

IMs encode COPs as the minimization of an energy landscape defined by an Ising Hamiltonian $H(\boldsymbol{m}) = -0.5 \sum_{ij} J_{ij} m_i m_j - \sum_i h_i m_i$ [22,25], where $\boldsymbol{J}$ is the coupling matrix, $\boldsymbol{h}$ is the bias vector, and $\boldsymbol{m}$ is the Ising vector. Our architecture implements the update rule of each p-bit, $m_i(t+1) = \operatorname{sgn}(\operatorname{rand}(-1,+1) + \tanh(I_i(t))$, where the local field term $I_i(t) =$ $\beta(t)I_{0_i}(t) = \beta(t)\left(\sum_j J_{ij}m_j(t) + h_i\right)$ is representing the collective behavior of the Ising spin vector[25].

The architecture has been validated using portfolio optimization and LDPC decoding. Figure 1a shows the block diagram of portfolio optimization problem, a central task in finance that is particularly challenging due to its full connectivity[31,35], as each asset interacts with all others through pairwise correlations. The objective is to determine the allocation vector $w$ for $M$ assets, where $w_i$ represents the fraction of total budget invested in $i$-th asset. The expected portfolio returns is given by $w^T u$, where $u$ is the vector of mean historical asset returns, while the portfolio risk is expressed as $w^T S w$, with $S$ is the covariance matrix encoding pairwise asset dependencies. The optimization problem consists of maximizing expected return while minimizing risk, subject to the constraint $\sum_i w_i = 1$, with ensures full allocation of the available budget. Further details on the mapping of portfolio optimization onto the Ising model are provided in the Methods section.

Figure 1b shows the block diagram of LDPC chain used in telecommunication, including encoder, bit-to-symbol mapper, noisy channel ($N$) and decoder. LDPC are linear block codes defined by a sparse parity-check matrix $P$ (or, equivalently, a generator matrix $G$) and are naturally represented as Tanner graphs, with variable and check nodes determined by the nonzeros of $H$. At the receiver, decoding seeks a codeword $C'$ that (i) satisfies all parity checks and (ii) remains close to the noisy observation $R$. This decoding task can be seen as a COP, combining a parity-consistency term and a data-fidelity term. The mapping of LDPC to the Ising model is described in detail in the Methods section.

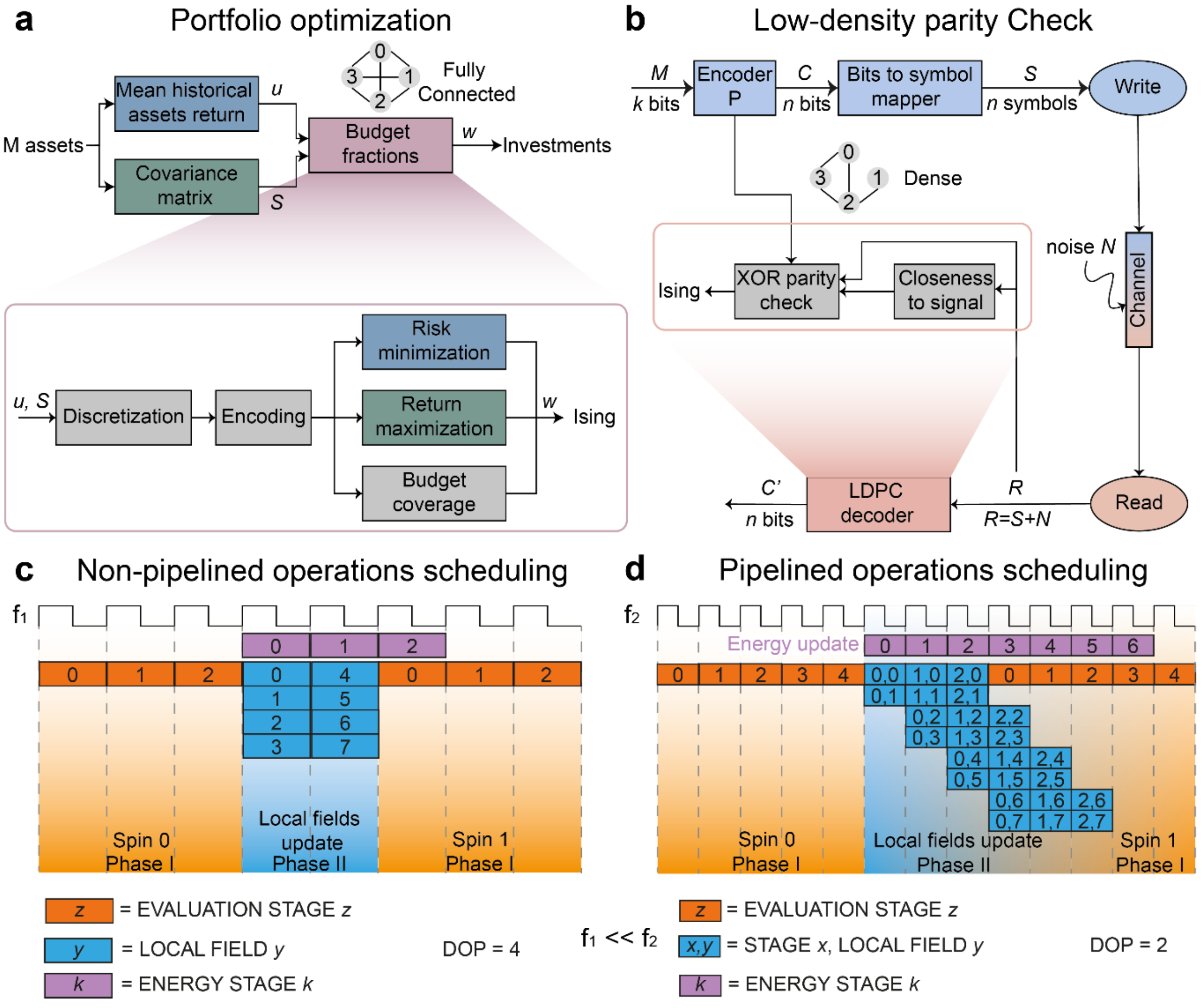

**Figure 1: a** Block diagram of the portfolio optimization problem: given *M* assets, the goal is to determine the optimal weights *w* that maximize the expected return (based on the mean historical assets return *u*), while minimizing the risk, quantified by the assets covariance matrix *S*. **b** Block diagram of the LDPC communication chain, including encoder, bit-to-symbol mapper, noisy read channel, and decoder. The decoding task is mapped to an Ising Hamiltonian. **c**-**d** Schematics of the PIM operations scheduling in a non-pipelined (**c**) and the proposed pipelined (**d**) architecture. In (**c**), the local-field and energy updates for all p-bits and the spin evaluations are executed sequentially; in (**d**), they are overlapped across p-bits, enabling concurrent execution, higher update rates and reduced hardware utilization. In (**d**) each blue block is labelled (x, y), where y is the index of the local field being updated and x is the stage of the update path processing it; in (**c**) each update completes within a single clock cycle and only the index y is required. Vertically adjacent blue blocks are executed in parallel, horizontally adjacent blocks are successive stages of the same update, and the horizontal offset between consecutive rows is the pipeline delay. On the AMD ZCU106 board, the clock frequencies are $f_1$ = 40 MHz and $f_2$ = 300 MHz. The two panels span the same time interval, so that the pipelined architecture sustains a comparable throughput at a lower DOP. The panels are schematic and not to scale: the ratio between the two clock periods and the number of pipeline stages shown are reduced for legibility, and no quantitative conclusion should be drawn from the cycle counts depicted.

## Results

As a first step, the quality of sampling of the proposed architecture has been compared with a software PIM on invertible logics (see Supplementary Note 1 and Supplementary Fig. 1) and instances of Max-Cut (Supplementary Note 2 and Supplementary Fig. 2). Secondly, we provided a detailed scalability and complexity analysis, quantifying logic utilization and memory requirements, achievable update rate as a function of problem size, and trade-offs among numerical precision, memory organization, and throughput. According to the available hardware on an AMD ZCU106 board, the proposed architecture can be synthesized to manage up to 512/1024/2048 p-bits using 16/10/2-bit fixed-point coefficient for for $\boldsymbol{J}$ and $\boldsymbol{h}$.

Throughout this section, and in the figures and Supplementary Information, we distinguish the following levels of the design and validation flow:

- *Implemented*: synthesis and place-and-route completed on the target device with timing closure met. Resource utilization and maximum clock frequency are taken from the Vivado™ implementation reports.
- *Measured / experimentally validated*: the FPGA was programmed with the generated bitstream and executed on the target board. Functional outputs and execution times were obtained from board-level measurements.
- *Estimated*: obtained from Vivado™ reports rather than from board-level runs. In the revised manuscript this term applies only to power.
- *Projected*: extrapolated from the scaling trends of the experimentally validated designs. These configurations were not implemented.

The validation level of every configuration reported in this work is summarized in Supplementary Table S2.

**A) Not pipelined architecture**

The baseline non-pipelined architecture implements a fully connected PIM and follows the operation scheduling shown in Fig. 1c. In this architecture, all couplings (elements of $\boldsymbol{J}$), local biases (elements of $\boldsymbol{h}$), and spin states $\boldsymbol{m}$ are stored on the FPGA using Block Random Access Memory (BRAM). To avoid recomputing full sums at every update, the local fields are first initialized and then updated whenever a spin $m_i$ flips, according to $I_{0_j} \leftarrow I_{0_j} + 2m_iJ_{ij}$. The architecture operates in a two-phase loop for each p-bit. The first phase evaluates whether the p-bit flips by comparing a random number in $[-1, +1]$ with the hyperbolic tangent of the local input $I_i$, given by the local field and the evolution factor $\beta$, and performs the flip if necessary. The second phase updates the local fields and the energy of the Hamiltonian. Local-field initialization and energy accumulation are handled by a multiplier-free Multiply-And-Accumulate (MAC) block (see Supplementary Note 3). The hyperbolic tangent is computed using an optimized approach combining linearization, a lookup-table (LUT) and saturation (see Supplementary Note 4), and the random numbers are generated with a xoshiro128++[33] (see Supplementary Note 5).

The update of the local fields $I_{0_j}$ can be computed in parallel for all or a subset of p-bits, according to the available hardware resources, as shown in Fig. 1c. The DOP, defined as the number of J-elements read per clock cycle (i.e. the number of local-field updates that can start concurrently), is configurable in our design. The evaluation of the next p-bit begins only after all local field updates have been completed. The scalability of parallel updates is limited for two main reasons: (i) the increased amount for hardware resources; (ii) the complex memory access patterns to manage. The latter arises because FPGA memories have a limited number of ports, restricting the number of elements that can be read per cycle. To address this, we adopted three standard techniques, word packing, paging, and multi-pumping, to virtually increase the effective memory bandwidth (see Supplementary Note 6). However, these techniques must be applied carefully; overly aggressive use tends to fragment the memory into scattered registers, wasting BRAM resources, complicating address generation, and ultimately limiting the scalability of the architecture (see Supplementary Note 6)[36]. Even with parallelization, the main bottleneck of this architecture remains the local fields update phase, labeled "Update MAC" in Figure 1c. This step quickly becomes the dominant contributor to both latency and computational cost as the problem size grows, unless the DOP is increased accordingly (see Supplementary Note 7A). In this architecture, the evaluation of p-bit updates and energy computation constitutes the critical combinatorial path, which limits the maximum achievable clock frequency to 40 MHz, a value comparable to previous fully digital implementations[19,25,37,38].

## B) Resource efficient pipelined architecture

Figure 1d shows the schematic representation of the proposed resource-efficient pipelined implementation of a PIM. It is designed to minimize the critical combinatorial path, overlap the two phases (evaluation and local field update) and optimize the operations scheduling in order to improve scalability and performance in terms of TTS and area occupancy. The pipeline depth is obtained by partitioning the combinational critical path of the non-pipelined architecture, which runs through the local-field adder, the multiplier, the tanh evaluation and the comparator, and which limits that architecture to 40 MHz, into stages, until the resulting stage delay meets the target frequency. The depth is therefore a consequence of the structure of that path and of the bit width of the data travelling on it, and not of the number of p-bits, which enlarges the coefficient memories while leaving the arithmetic unchanged, nor of the DOP, which replicates the update datapath without lengthening any of its stages. The 23 stages reported here are the minimum depth at which this path meets the 300 MHz constraint on the AMD ZCU106 at 16-bit coefficient and 32-bit accumulation width. The stages are placed inside the arithmetic blocks, so that changing the depth requires re-partitioning and re-verifying them rather than re-elaborating the same description at a different setting: the depth is a design decision taken once, for a given arithmetic precision and target frequency. In this design, the local field update sequence is reordered to prioritize the update of p-bit $i + 1$, such arithmetic reordering does not alter the logical ordering of the p-bit state update. This removes the critical data dependency between consecutive evaluations, allowing the next p-bit evaluation to start as soon as the local field of p-bit i+1 has been updated, while the remaining local-field and energy updates proceed in the background. Thanks to the pipelined organization, a new local field update is launched at every clock cycle, guaranteeing that the pipeline remains continuously filled. The complete schedule of both architectures is reported as pseudocode in Supplementary Note 7. In the pipelined design a new p-bit evaluation is issued every $L_{ev} + L_{up}$ cycles, against $L_{ev} + N/\mathrm{DOP}$ for the non-pipelined baseline, where $L_{ev}$ and $L_{up}$ are the depths of the evaluation and update paths: this difference is the origin of the throughput advantage. By carefully scheduling memory accesses, the proposed architecture preserves strictly sequential-update sampling for any DOP. A moderate DOP with respect to the number of p-bits is sufficient to keep the pipeline filled and the latency almost constant as the problem size grows, without any impact on the sampling order. As a result, the design achieves up to an order-of-magnitude speedup over the non-pipelined baseline at fixed DOP, without requiring full replication of the update datapath across all spins.

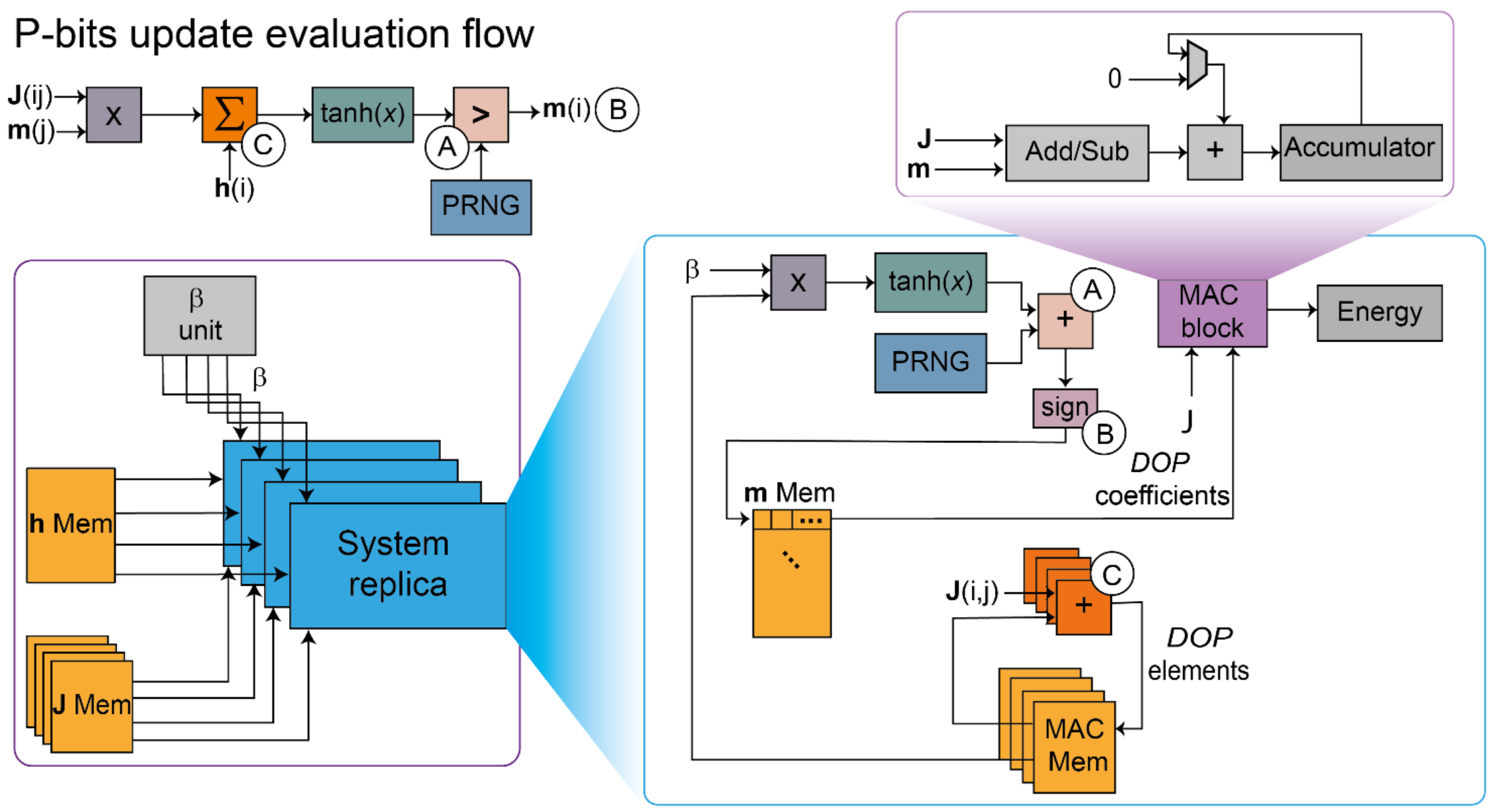

**Figure 2**: Hierarchical view of the pipelined FPGA-based PIM. The p-bits update evaluation flow is reported on the top of the figure, with each step mapped to the corresponding datapath components. The left side shows multiple system replicas sharing coupling and local-bias memories, while the right side details the internal structure of a single replica. Each replica includes a pseudo-random number generator (PRNG) (xoshiro128++), a hyperbolic tangent unit (tanh), on-chip memories for local fields (MAC Mem) and p-bits states (**m** Mem) storage, a multiplier-free MAC, dedicated adders for local-field updates. The $J_{ij}$ are read from on-chip memory to initialize the local field and the energy of the current spin configuration via a MAC tree, and to update local fields upon spin flips. Memories are paged/banked to sustain continuous pipeline throughput.

Figure 2 displays the detailed hierarchical organization of the pipelined PIM associating the datapath components with the steps of the p-bit update evaluation flow (top of the Fig. 2). It allows the concurrent execution of multiple independent replicas of the same system, each evolving under identical parameters but with independent stochastic dynamics, as illustrated on the left-hand side of the figure. These replicas share the same $\boldsymbol{J}$ and $\boldsymbol{h}$ memories while maintaining private state memories for their spin configurations and local fields (MAC values), enabling more sophisticated annealing algorithms such as parallel tempering and simulated quantum annealing[22]. We also wish to highlight here that this pipelined architecture can be used for momentum annealing with a minor modification (not shown in this paper) allowing parallel p-bit updates[39]. The number of independent parallel replicas is only limited by the available resources.

Within a single replica, the key components include a pseudo-random number generator (xoshiro128++), a hyperbolic tangent unit, on-chip memories for storing the precomputed local fields (MAC Mem) and the p-bit states ($\boldsymbol{m}$ Mem), a multiplier for including the evolution of the inverse-pseudo temperature parameter $\beta$ into the local-field evaluation, and a multiplier-free MAC unit (Supplementary Note 3) for local-field initialization and energy updates. The p-bit flip decision is implemented by a comparator consisting of an adder (highlighted as ($A$) in Fig. 2) and a sign detector (indicated as ($B$)), while dedicated adders handle local-field updates (highlighted as ($C$)). All modules are coordinates by a centralized control unit.

At the initialization step, each module accesses the coupling coefficients $J_{ij}$ from shared memory to compute the initial values of the local field and replica energy. Subsequent memory accesses occur during MAC operations and energy updates whenever a p-bit changes state. To maintain high throughput, word packing and multi-bank memory organization are employed, allowing each memory access to return a vector of coefficients aligned with MAC tree and the local-field update adders (see block ($C$)).

The hyperbolic tangent function, used to evaluate the p-bit flip probability, is implemented through a combination of linearization, LUTs, and saturation (see Supplementary Note 4), while the memories are paged and banked to support continuous pipeline operation. The output of the hyperbolic tangent unit is compared with the generated random number (block ($A$)), and the p-bit flip is accepted or rejected based on the outcome of the sign detector (block ($B$)).

The memory bandwidth-aware organization, combined with the pipeline structure and optimized operations scheduling reduces the critical path and increases the maximum operating frequency. In this design, logic utilization grows primarily with the chosen DOP that is, with the number of local field updates computed in parallel, rather than linearly with the problem size.

The results presented here consider problem instances in which all couplings $\boldsymbol{J}$, local biases $\boldsymbol{h}$, and spin states $\boldsymbol{m}$ are stored on the FPGA memory. This configuration removes the off-chip bandwidth limitations and natively supports dense connectivity. Nevertheless, the proposed pipelined architecture is scalable to multiple FPGA PIM architecture, compatible with the use of off-chip memory, and with the use of extended probabilistic variables such as p-dit[40] with minor modifications.

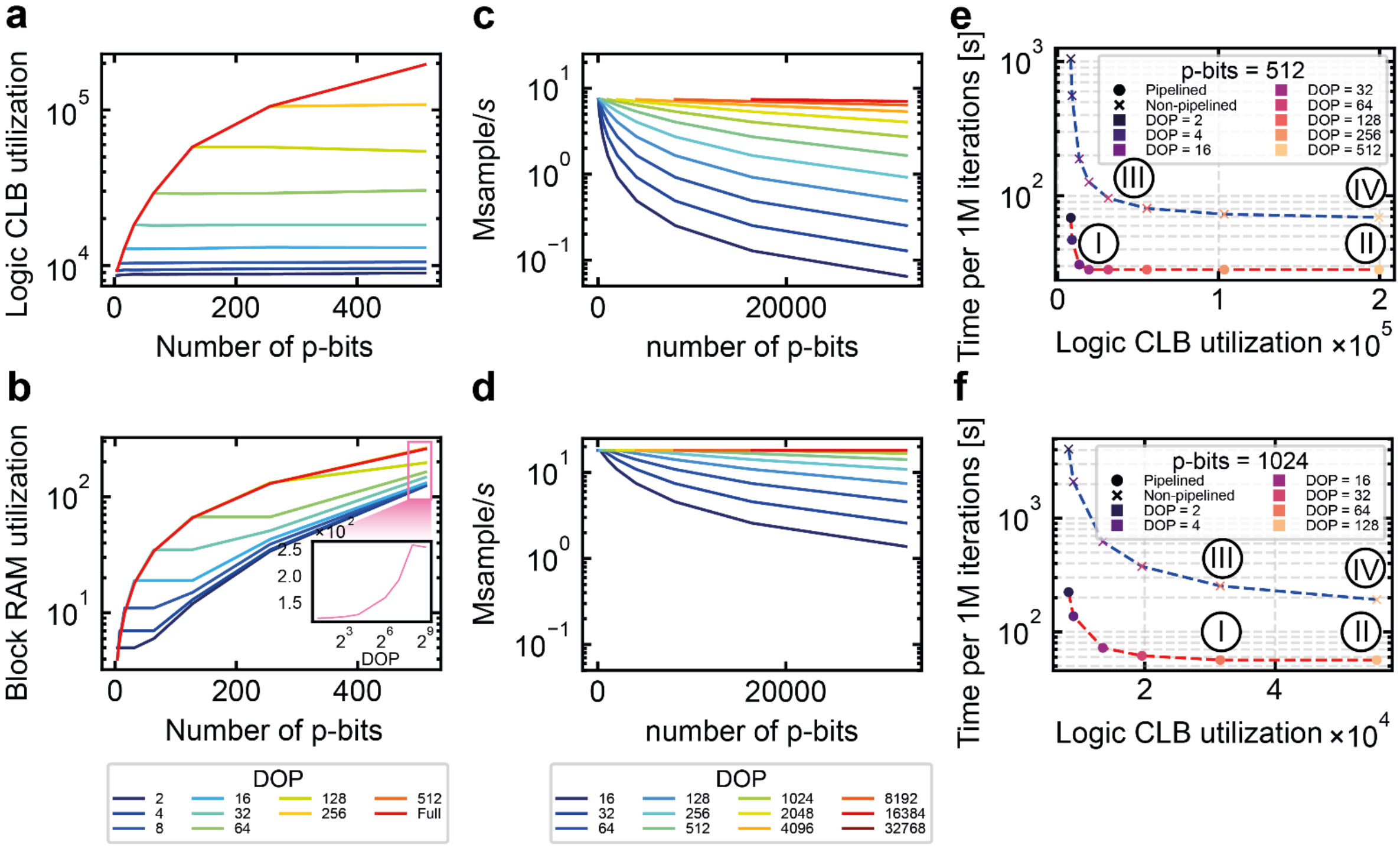

**Figure 3**: **Throughput and resource scaling on an AMD/Xilinx ZCU106 a** Logic CLB utilization as a function of the number of p-bits for different DOP (16 bit data parallelism for Ising model representation). **b** Same as **a** but for Block RAM tile utilization (16 bit data parallelism). **c** Update throughput (Msamples s⁻¹) versus number of synthesized p-bits for the non-pipelined design (40 MHz clock frequency). Results above 512 p-bits are a projection; the 512-p-bit designs are experimentally validated (16 bit data parallelism). . **d** Update throughput (Msamples s⁻¹) versus number of synthesized p-bits for pipelined architecture with optimal pipeline depth (23 pipeline levels in the evaluation and local field update phase, and 300 MHz clock frequency). Results above 512 p-bits are a projection; the 512-p-bit designs are experimentally validated (16 bit data parallelism). **e** Comparison of the time required for 1 million of simulated annealing iterations as a function of logic CLB utilization on the FPGA with an architecture for 512-p-bit, considering 16 bits for representing J elements for pipelined and non-pipelined designs (DOP from 2 to 512). Points I–IV highlight the execution time and CLB utilization for the pipelined and non-pipelined designs, considering DOP values of 32 and 512 **f** same as **e** with an architecture for 1024-p-bit, considering 8 bits for representing J elements (DOP from 2 to 128, the largest DOP that the device accommodates at this size and precision). Points I–IV highlight the execution time and CLB utilization for the pipelined and non-pipelined designs, considering DOP values of 64 and 128.

## C) FPGA implementation and synthesis results

The designed PIM architecture is synthesized on an AMD Xilinx ZCU106 to validate the timing and functionalities. The clock frequency reached with 23 pipeline stages is 300 MHz, that is one order of magnitude larger than the 40 MHz achieved by the non-pipelined design for the same logic and memory utilization. The PIM is synthesized for different DOP. The notation, DOP = $K$ indicates that the local field updates are performed for $K$ p-bits in parallel, while Full DOP denotes the configuration in which the parallelization degree is equal to the total number of synthesized p-bits.

Figure 3a shows how the number of Configurable Logic Blocks (CLBs) scales as a function of the p-bits number for different DOP. For a fixed DOP, the number of CLBs used on the FPGA is almost independent of the number of p-bits. This is because the MAC width and the number of adders required for local field updates scale with the DOP and not with the number of p-bits, while the area occupancy of the other elements, such as the hyperbolic tangent unit, are independent of both. On the other hand, the Full DOP case (red line in Fig. 3a) describes the number of CLB as a function of the p-bit with a DOP equal to the number of p-bit, it exhibits a rapid increase because scaling the DOP proportionally to the size of the Ising spin vector significantly increases the MAC tree size and linearly increases the number of adders required for local field updates, which quickly becomes the dominant contribution to the logic area occupancy.

Figure 3b shows the BRAM utilization as a function of the number of p-bits for different DOP. It is possible to notice that the BRAM usage grows at least quadratically with the number of p-bits, as storing and streaming the coupling matrix $\boldsymbol{J}$ is the dominant contribution in the memory requirements. While BRAM usage is expected to remain roughly constant when increasing the DOP for a fixed number of p-bits, a slight increase is observed due to less efficient memory tile allocation when paging and multi-row banking are used to satisfy higher bandwidth requirements. This effect can be better observed in the inset of Fig. 3b where the BRAM utilization as a function of the DOP is reported for the 512 p-bits configuration. Therefore, the Full DOP design (red line) represents the upper limit of BRAM occupation all the number of p-bits.

Figures 3c and 3d show the p-bits update rate (Msamples/s) as a function of the number of synthesized p-bits at different DOPs, for the non-pipelined and pipelined architectures, respectively. For DOP ≥ 1024 p-bits, the curves include projected values. In the non-pipelined design, the throughput decreases approximately proportionally to the ratio between the number of p-bits and the DOP, because each update must wait for the full local-field and energy update phase to complete.

Consequently, the update rate rapidly degrades as the problem size increases, even when large DOPs are used. On the other hand, the pipelined architecture overlaps the evaluation and update phases, which significantly reduces the throughput degradation associated with larger p-bit counts and avoids the need for proportionally increasing the DOP. While increasing DOP improves performance in both architectures, the pipelined design preserves these benefits across growing sizes, since its logic area scales primarily with DOP and not with the problem dimension. At fixed DOP, the pipelined PIM achieves a speed-up exceeding one order of magnitude compared to the non-pipelined baseline (Supplementary Note 8), confirming pipelining as a key architectural strategy for scalable, high-throughput PIM implementations. All quantitative comparisons between the pipelined and the non-pipelined architectures reported in this work are performed at fixed DOP. Three comparison protocols are used throughout this work, and each reported figure refers to one of them. Under fixed DOP, both architectures are implemented with the same DOP, which isolates the effect of pipelining from that of parallelism; this protocol underlies Figs. 3c–d, the operating points I–IV of Figs. 3e–f, and Fig. 4. Under fixed resource budget, the two are compared at equal logic utilization. Since for a given problem size and DOP the two implementations occupy essentially the same logic (the differences arising from the additional pipeline registers and control logic rather than from any replication of the arithmetic datapath) this protocol coincides with the previous one, and Figs. 3e–f report the execution time as a function of the occupied CLBs so that the performance–resource trade-off can be read directly. Under individually optimized settings, each architecture is assigned the DOP optimal for it: for the pipelined design this is the balanced operating point discussed in Supplementary Note 7, whereas the non-pipelined design benefits from parallelism up to full DOP. Comparing each architecture at its own optimum is what the throughput and resource figures in the following sections refer to when a single best-case value is reported.

Figure 3e reports the time required to perform one million simulated-annealing iterations for solving an optimization problem mapped onto a 512-p-bit architecture, as a function of the FPGA logic area measured in CLBs, for different DOP and for both the non-pipelined and pipelined designs. It can be observed that increasing the DOP initially reduces the execution time for both the non-pipelined and pipelined architectures, at the cost of an approximately linear increase in logic-area utilization (results are reported for power-of-two scaling values of DOP). Beyond a certain DOP, the pipeline becomes fully occupied. In this regime, further increases in DOP lead to larger area occupancy without providing additional throughput benefits because of stalls occurrences. In such a condition, the execution time saturates and approaches a horizontal asymptote at approximately 40 s, as indicated by points I) and II) in Fig. 3e. Along the DOP axis, increasing the parallelism raises the logic utilization while improving the throughput up to the balanced operating point discussed in Supplementary Note 7. At N = 512 the logic utilization ranges from 18,225 LUTs to 196,534 LUTs across the DOP range, confirming that the pipelined architecture attains its throughput at a small fraction of the resources required by full parallelism. On the other hand, an increasing of the DOP in the non-pipelined architecture give rise to a continuous improvement in the execution time because as expected no saturation is observed even at high DOP values (compare point III with IV in Fig. 3e). However, the pipelined design obtains a given performance level with significantly lower hardware cost thanks to the higher operating frequency and the improved operation scheduling enabled by overlapping the evaluation and local-field update phases. Overall, these results highlight the key advantage of the pipelined approach in terms of the timing-area trade-off. In particular, the proposed design achieves nearly one order of magnitude speed-up over the non-pipelined architecture while reducing the logic CLB utilization by eight times.

Overall, the results demonstrate that a strategy based on the use of a moderate DOP with respect to the number of p-bits (formal demonstration in Supplementary Note 7B), reduces both the logic and memory utilization enabling high performance through pipelining implementation. A comparison with state-of-the-art PIM architectures is presented in Discussion section.

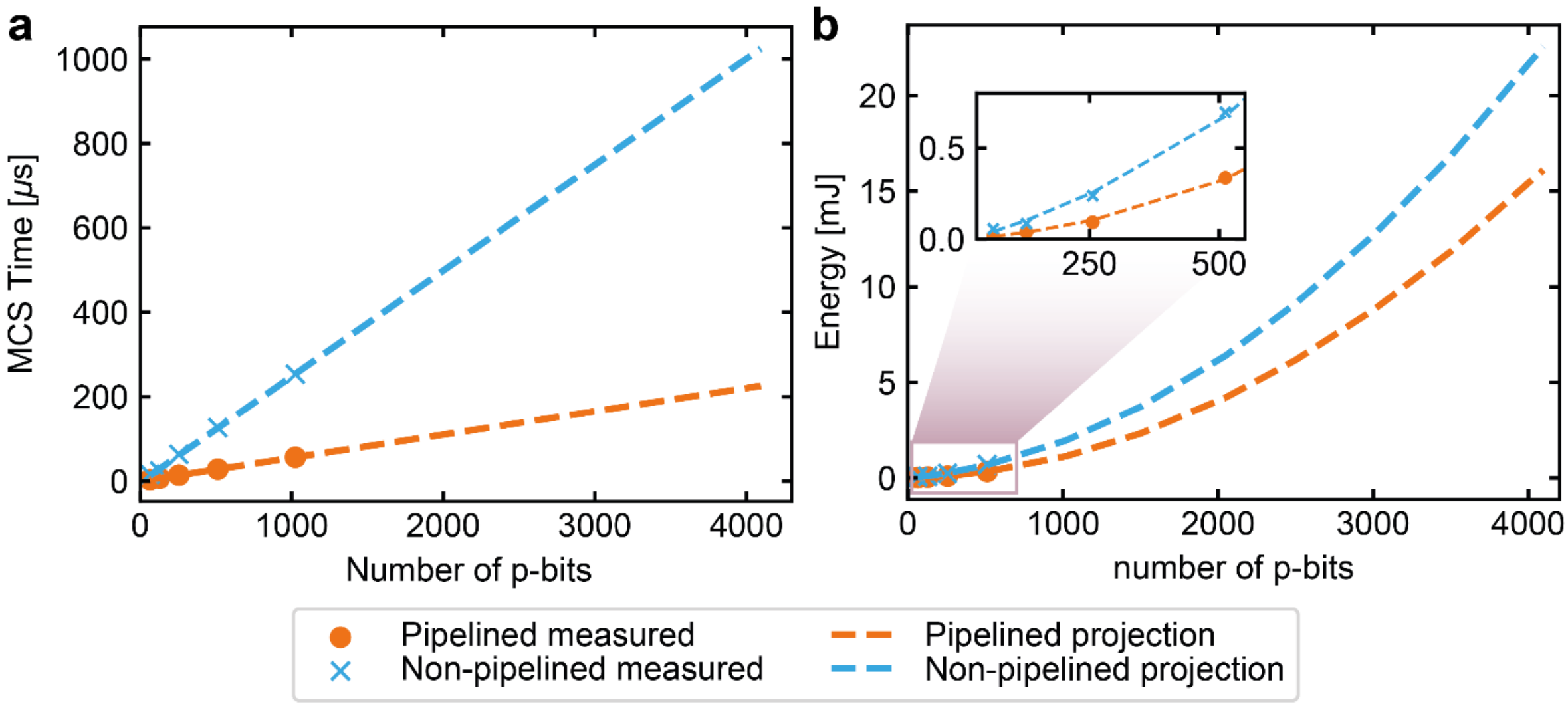

**Figure 4**: **Architectural performance**. **a** Measured and projected time required to perform a complete Monte Carlo Sweep (MCS) for pipelined and non-pipelined architectures as a function of the number of p-bits. Results assume an 8-bit representation for the coefficients and the optimal DOP for the pipelined architecture. **b** Estimated and projected energy required to perform a complete MCS for the same architectures, obtained as the product of the total power reported by VivadoTM (static plus dynamic) and the measured MCS time of panel a.

**Hardware performance for portfolio optimization and LDPC problems.**

We have evaluated the performance of the PIM on several portfolio optimization instances derived from publicly available financial datasets[41] and on a suite of LDPC instances generated with the Pyldpc library[42] and cast their decoding into the Ising cost function.

In portfolio optimization, for each asset, historical return and covariance information was extracted and used to construct the corresponding Ising model, following standard mean-variance formulations (see method section). We systematically varied both the number of assets included in the portfolio and the numerical precision used to encode the decision variables, generating families of problems with increasing dimensionality and encoding granularity. This methodology enables a controlled assessment of scalability, numerical robustness, and hardware–software consistency across a wide range of realistic optimization scenarios.

On the other hand, a suite of LDPC instances was generated by varying the code length, the Signal-To-Noise Ratio (SNR), and the sparsity of the parity-check matrix. Performance was evaluated in terms of Bit Error Rate (BER), defined as the fraction of bits decoded incorrectly, and the TTS ).

Since the proposed PIM is fully-connected, the execution time and energy required to perform a Monte Carlo Sweep (MCS), i.e. a simulated annealing iteration, depend only on the number of p-bits and the chosen DOP. Therefore, both metrics are totally independent of the specific optimization problem. Consequently, the architectural scaling presented in previous section applies equally to both the portfolio optimization and LDPC benchmarks considered in this work.

Figure 4 reports the measured and projected architectural performance of the system observed with both the benchmark considering a single replica of the system. In Fig. 4a, the time required to perform a complete MCS is shown as a function of the number of p-bits for both pipelined and non-pipelined implementations, considering the optimal DOP for the pipelined architecture (about 30% of the number of p-bits). Measured results are reported for designs up to 1024 p-bits, assuming 8 bits representation for the ***J*** matrix coefficients, while projections for larger systems are obtained from the observed scaling trends. The pipelined architecture shows a 5x lower execution time compared to the non-pipelined design.

Figure 4b reports the corresponding dynamic energy required to perform a complete MCS, computed as the product of the total power estimated by Vivado™, static plus dynamic, and the measured execution time. The results show that the pipelined architecture also provides a 52% energy consumption reduction on experimentally validated designs and about 29% on projections, as the reduced execution time more than compensates for the increase in dynamic power consumption related to the higher operating frequency of the architecture. The reported energy accounts for the annealing kernel only. The transfer of the $J$ matrix and the $h$ vector into on-chip memory occurs once before the run and is never repeated during execution (Supplementary Note 6); it is identical for the two architectures and therefore does not affect the comparison. As a result, the pipelined design achieves both improved time-to-solution for a fixed number of MCS and lower energy per sweep.

A more detailed results obtained on these benchmark instances are reported in Supplementary Note 9. Looking ahead, several directions could further enhance the hardware performance of this architecture. On-chip caching and reuse of frequently accessed coupling submatrices or partial local field accumulators may alleviate BRAM pressure and reduce memory bandwidth demands[43]. Advanced annealing strategies, including parallel tempering, and simulated quantum annealing[22] can be integrated. Finally, an ASIC implementation could exploit the further advantage of fine-grained SRAM banking, clock-gating, and near-memory computation[44], offering a direction substantial energy efficiency and speed improvements beyond current FPGA limits.

**Discussion and comparison with state-of-the-art PIM architectures**

State of the art PIM architectures (see Table 1 for a comparison), fall into three classes: (i) low-latency but limited-scale; (ii) scalable but throughput-limited; or (iii) fast and scalable only under sparsity assumptions (time-multiplexed/sparse couplings, not general for dense fully-connected cases). Early works on PIMs[19] proved modularity and proof-of-concept for reversible logic but storing of $\boldsymbol{J}$ and $\boldsymbol{h}$ (low-precision coefficients, only 6 bits) were based on local registers. Massively parallel architectures[25] benchmarked in integer factorization, and several COPs showed that flips-per-second scale linearly with the number of p-bits with a crucial dependence on sparsification of the coupling matrix and external Double Data Rate Fourth Generation (DDR4) memories that results in low operating frequencies (15–30 MHz) and no direct support for fully-connected graphs. More recently, the architecture pc-COP[37] achieved up to 2048 p-bits fully-connected on-chip using a speculate-and-select pseudo-parallel scheme, with high accuracy on Max-Cut benchmarks. Nevertheless, limiting the coupling representation to two bits severely limits the problems that can be solved confining the architecture to Max-Cut formulations with {-1, 0, 1} weights. This constraint, combined with the overhead of speculate-and-select parallelism, limits generality, achievable clock frequency, and scalability when extending to more complex classes of weighted Ising problems.

In[38], the authors prove an impressive sweep-time $O(1)$ scaling on sparse, structured 3D lattices by exploiting graph-coloring and replica-based APT/ICM and projecting large replica counts on ASICs and stochastic Magnetic Tunnel Junction (sMTJ) devices. However, the presented FPGA realization fits only one large replica at a time, so multi-replica APT/ICM is sustained via time-division multiplexing with off-chip CPU handling of swaps and cluster moves, and several conclusions rely on assumptions that all replicas run in parallel on a single chip.

While previous FPGA-based PIM architectures are well suited for sparse Ising graphs through replica parallelism, they are fundamentally limited by memory bandwidth and on-chip data movement when extended to dense, fully-connected problems. On the other hand, our architecture directly targets this bottleneck by providing a fully on-chip, pipelined all-to-all update path with bandwidth-aware memories and a multiplier-free multi-operand MAC, avoiding external DDR access. Compared with prior FPGA-based PIM architectures, the proposed design simultaneously achieves the highest reported operating frequency, higher-precision coupling coefficients, and scalable support for dense fully-connected Ising models without relying on sparsification.

The results reported for the AMD Alveo U250 correspond to completed synthesis and place-and-route with timing closure met, including the 300 MHz operating frequency verified after place-and-route on that device; no board-level execution was performed on the AMD Alveo U250.

| Design | Maximum Size | Clock Frequency | Data Parallelism | Pros | Cons |
|---|---|---|---|---|---|
| [19] | 448 p-bits Without external memories (on AMD Kintex ultrascale) | 100 MHz | 6 bits (32 bits only for PRNG) fixed point | Modularity, accurate proof of concepts for reversible logic | Local registers for $J$ and $h$ storage, LFSR for PRNG, full LUT implementation for tanh, no total energy estimation |
| [25] | 4793 p-bits using external DDR4 memories (on AMD VCU118 UltraScale+) | 15–30 MHz (using 5 clock phases at 15 MHz or 4 phases at 30 MHz) | 10 bits (32 bits only for PRNG) fixed point | Massively parallel, flip rate scales linearly with N Demonstrates applications such as 32-bit integer factorization and approximate SAT solving. Architecture is scalable with sparsification. | Requires exact sparsification of the problem. Cannot directly handle fully-connected graphs. Low operating frequency (15–30 MHz). More complex implementation, strongly dependent on problem sparsity. |
| [37] | 2048 p-bits (on AMD ZCU104) without external memories | 100 MHz | 2 bits (32 bits only for PRNG) fixed point | Fully-connected pseudo-parallel update (speculate-and-select), accuracy of ~99% on G-Set. | Specialized to Max-Cut with 2-bit weights, simplified activation and costly speculate-and-select parallelism, limiting generality, scalability, and frequency. |
| [38] | 4096 p-bits | 40 MHz | 10 bits (32 bits only for | Uses fully on-FPGA BRAM for $J$, $h$ and states, | On FPGA only one large-scale replica fits at once, so |

| | (on FPGA AMD Alveo U250) | | PRNG) fixed point | achieves high throughput via graph coloring with strong energy efficiency, and scales to large 3D Ising instances through time-division multiplexing of replicas | multi-replica methods require time-division multiplexing and off-chip CPU handling for swap/ICM, with capacity bounded by BRAM. |
|---|---|---|---|---|---|
| This work (Non Pipelined) | 512/1024/2048 p-bits (on a AMD ZCU106) without using external memories | 40 MHz | 16/10/2 bits (32 bits only for PRNG) fixed point | Uses fully on-FPGA BRAM for $J$ and $h$. The parallelization degree for local field update and the data parallelism optimize according with needs and available resources. | To reduce the latency, it is necessary to increase parallelization, and, consequently the bandwidth of the memory with negative impact on memory and resource allocation on FPGA. |
| This work (Pipelined) | 512/1024/2048 (AMD ZCU106) and 2048/4096/8192 (AMD Alveo U250) p-bits without using external memories | 300 MHz | 16/10/2 bits (32 bits only for PRNG) fixed point | Pipelining maintains high throughput at reduced parallelism, lowering resource costs without relying on sparsification. | If the degree of parallelism and pipeline depth are not properly tuned, stalls occur and peak throughput is not achieved. |

**Table 1**: Comparison of FPGA-based probabilistic Ising machine architectures reported in the literature. For each design, the maximum problem size, operating frequency, numeric precision, validation level, and main pros/cons are highlighted. The validation level distinguishes configurations that were experimentally validated on the target board from those for which synthesis and place-and-route were completed but no board-level execution was performed.

A recent theoretical study on pipelined multi-operand adders[45] also targeted fully-connected updates. It represents a useful conceptual precedent on pipelined adders for fully-connected p-bits networks, but it does not prove scaling, and the reported throughput is based on simplified assumptions (ideal stall-free pipeline, neglecting the tanh evaluation, and abstract gate-delay estimates). Moreover, the pipelining strategy differs from the present work, as it is based on re-computing the input signal “on the fly” at each update rather than overlapping state refresh and flip evaluation. Our proposed pipelined approach can also be adapted to probabilistic algorithms that are based on the parallel update of p-bits, such as momentum annealing[39]. In this context, our contribution is complementary rather than competitive. In fact, while momentum annealing is an algorithmic technique designed to accelerate sampling dynamics through inertial terms, our work focuses on the hardware realization of a scalable and resource-efficient update engine that can support such algorithms without requiring full hardware duplication as the number of p-bits increases. In particular, the proposed strategy allows the designer to specify the maximum number of p-bits that may initiate an update in parallel, i.e., the degree of hardware replication, while additional p-bits are processed concurrently through the pipeline stages, preserving high throughput with bounded resource growth. At the same time, the proposed pipelined architecture can be extended to support higher-order Ising models, like in[46], by restructuring the coefficient storage and accumulation strategy. Rather than associating interaction coefficients with individual p-bits, higher-order terms can be naturally accommodated by organizing coefficient memories around interaction hyperedges, with the corresponding contributions streamed through the same multi-operand accumulation pipeline. This approach preserves the core design principles of the proposed update engine, bounded hardware replication, fully overlapped pipelined execution, and bandwidth-aware on-chip data movement, while enabling the evaluation of multi-body interactions without requiring fundamental changes to the control flow. As a result, the architecture retains scalability and throughput even when moving beyond pairwise couplings, making it well suited for probabilistic models that rely on higher-order interactions.

The architectures compared are benchmarked on different problem classes and at different coefficient precisions. A p-bit update does not represent the same amount of arithmetic across the table, so throughput and energy per sweep are to be read together with the precision column. The update primitive itself differs: some of the compared designs require exact sparsification of the coupling matrix or exploit graph coloring to update an entire colour class at once, so that a sweep is not the sequential Gibbs sweep implemented here, and the corresponding throughput figures are obtained under structural assumptions on the problem graph that do not apply to dense, fully connected instances

## Summary and Conclusion

We have presented a pipelined FPGA implementation (>20 pipeline stages) of a fully connected PIM designed for dense optimization problems. By overlapping spin evaluation with local-field and energy update, and by co-designing the datapath with a bandwidth-aware on-chip memory organization, the architecture mitigates the update and memory bottlenecks that typically limit digital PIMs on dense graphs. On an AMD ZCU106, the pipelined design reaches 300 MHz and provides more than an order-of-magnitude higher update throughput than a non-pipelined implementation at fixed parallelism, while maintaining controlled logic growth with the DOP rather than with problem size.

Validation against software PIMs indicates that the fixed-point implementation preserves the solution quality, and the case studies on LDPC decoding and portfolio optimization show that the architecture improves the sampling time of 10x and reduce the energy consumption for representative dense workloads. These results establish pipelining as a practical route to scaling fully connected PIMs on reconfigurable hardware without relying on sparsity, graph embedding or off-chip memory in the critical path.

At the same time, the present study identifies clear directions for further development. Larger fully measured implementations, direct comparisons with application-standard solvers, and explicit energy measurements will be important for assessing system-level advantage. Beyond FPGA prototypes, the same architectural principles could be extended to ASIC realizations and to hybrid probabilistic computing platforms. Overall, this work provides a solid hardware foundation for dense probabilistic optimization engines and a clear path towards more scalable digital and hybrid Ising systems.

## Methods

### Portfolio optimization problem mapping.

To describe the portfolio optimization problem in a standard QUBO/Ising form it is necessary to expand the involved continuous variables into binary form through proper encoding mechanisms. In this work, logarithmic encoding is adopted to discretize the asset weights, enabling an efficient representation of a wide dynamic range of values with a limited number of binary variables[47] Specifically, each asset weight $w_i$ is described as:

$$w_i = \sum_{z=0}^{p} 2^{z-p} b_z \, ,$$

where $b_z$ are the binary decision variables $\in (0,1)$ and $p$ denotes the number of binary variables used to encode each weight. This representation allows $w_i$ to represent values in the interval [0,1] with a minimum resolution of $2^{-p}$, defining the smallest representable increment in the asset weight. By increasing the precision, finer granularity in the weight discretization can be achieved at the cost of a linear increase in the number of binary variables, enabling a flexible trade-off between solution accuracy and problem size.

The continuous mean-variance portfolio optimization problem for $M$ assets is formulated as

$$\min H(w) = -w^T u + q w^T S w, \quad \text{subject to the budget constraint } \textstyle\sum_i w_i = 1,$$

where $\boldsymbol{u}$ is the vector of mean historical asset returns, $\boldsymbol{S}$ the covariance matrix, and $q$ the risk-aversion weight that sets the trade-off between return maximization and risk minimization. The budget constraint is included into the objective function through a quadratic penalty term:

$$H(w) = -w^T u + q w^T S w + \lambda \left(\textstyle\sum_i w_i - 1\right)^2,$$

where λ is the penalty weight enforcing the budget constraint. Substituting the logarithmic encoding introduced above ($w_i = \sum_z c_z b_{iz}$ with $c_z = 2^{z-p}$, resulting in $N = M \times p$ binary variables) into the penalized objective and using the binary property $b^2 = b$, the problem is transformed into the QUBO form:

$$H(b) = \sum_{(i,z)\neq(j,y)} (qS_{ij} + \lambda)c_z c_y b_{i,z} b_{j,y} + \sum_{i,z} \left[(qS_{ij} + \lambda)c_z^2 - (u_i + 2\lambda)c_z\right] b_{iz} + \lambda,$$

in which the return term contributes the linear coefficients $-u_i c_z$, the risk term the quadratic coefficients $q\, S_{ij} c_z c_y$, and the budget penalty both the quadratic coefficients $\lambda c_z c_y$ and the linear coefficients $-2\lambda c_z$. The constant λ does not affect the minimization and can therefore be discarded. Writing the quadratic and linear coefficients above as $Q$ and $L$ respectively, and applying the transformation $b = (1 + m)/2$, the QUBO is mapped onto the Ising Hamiltonian:

$$J_{(i,z),(j,y)} = \frac{1}{2}(q\, s_{ij} + \lambda)c_z c_y, \qquad h_{i,z} = \frac{1}{2}\left[\sum_{(j,y)\neq(i,z)} Q_{(i,z),(j,y)} + L_{i,z}\right].$$

For all portfolio instances reported, the relative weight between return and risk was fixed to $q$ = 1. This parameter was not optimized because the objective was not to study the efficient frontier, but to validate the FPGA implementation against the software reference on a dense, industrially relevant workload. The penalty weight λ is not fixed a priori: it is estimated per by MQT QAO following the procedure of[48], which bounds the range of the unconstrained cost function and sets λ above that bound, so that no configuration violating the budget constraint can be energetically favored over a feasible one. The library performs the substitution.

**LDPC problem mapping.**

The LDPC problem can be written into QUBO/Ising form by mapping bits to spins and translating XOR parity constraints into quadratic interactions introducing auxiliary variables, allowing LDPC decoding to be solved as an Ising optimization problem[30]. Indeed, these additional auxiliary variables create density. Specifically, considering $\tilde{C}$ the vector of bipolar binary variables, the cost function is composed of two parts: the parity check satisfaction and the closeness to the received signal. The former can be written in terms of spin variables as

$$\mathcal{P}(m) = \sum_{j=0}^{m-1} \frac{1}{2}\left(1 - \prod_{H_{i,j}=1} m_i\right),$$

where $m_i$ is the $i^{th}$ spin variable of $\tilde{C}$. The latter can be written as

$$\mathcal{R}(m) = \sum_{i=0}^{n-1} (R_i - m_i)^2.$$

The final cost function obtained is

$$H = \mathcal{R}(m) + \alpha\mathcal{P}(m)$$

where the weight α balances the parity term against the received-signal term. The parity term $P(m)$ is a product over the spins connected to each parity-check node. For the LDPC codes considered in this work, generated with a variable-node degree $d_v = 2$ and a check-node degree $d_c = 3$, each check therefore contributes a cubic monomial, so that *H* is a higher-order (cubic) Ising model. The weight is set to $\alpha = 5$ for all instances reported.

The cubic model is converted into quadratic form using the qubovert library, which applies a substitution (conjunction-gadget) reduction. Each cubic monomial $m_i m_j m_k$ is rewritten by introducing an auxiliary variable *z* representing the product $m_i m_j$ , so that $m_i m_j m_k \rightarrow z m_k$. The constraint $z = m_i m_j$ is enforced through a Rosenberg penalty term

$$\lambda\left(3z + m_i m_j - 2m_i z - 2m_j z\right),$$

which is zero when the constraint is satisfied and strictly positive otherwise. The pairs of variables to be substituted are selected by the library according to their frequency of occurrence across the higher-order terms, and the penalty weight $\lambda$ is set by the library default, which scales with the coefficient of the term being reduced.

The reduction introduces one auxiliary variable per parity-check node. An instance of code length $n$ has $m = \frac{nd_v}{d_c} = 2n/3$ check nodes, corresponding to a code rate of 1/3, and is therefore mapped onto $N = n + m = 5n/3$ p-bits. The code lengths reported in this work, $n = 300$ to 570 in steps of 30, thus correspond to $N$ = 500 to 950 p-bits. The horizontal axis of Supplementary Fig. 11 reports the code length $n$; the corresponding number of p-bits follows from this relation. The auxiliary variables introduced by the reduction are also what make the decoding instances dense, and hence a relevant workload for the proposed architecture.

The reduction reshapes the energy landscape, since it enlarges the search space from $n$ to $N$ variables and adds penalty terms with their own minima. The ground-state correspondence between the original higher-order formulation and the reduced quadratic one is nevertheless preserved, because the penalty weight is set so that violating a constraint always costs more than the energy it can save.

**Author contributions**

G.F. initiated the project and conceived the idea with inputs from P.K.A., H.Y. and M.Ch. G.F., M.Ch, M.Ca, and A.Gi secured the funding. D.V. designed the FPGA architecture and performed related testing and simulations. E.R. performed the software simulations. D.V., and G.F. wrote the manuscript with inputs from E.R. and A.Gr. All authors discussed the results, contributed to the data analysis, and commented on the manuscript. The study was performed under the leadership of G.F.

**Acknowledgements**

D.V., E.R, A. Grimaldi, A. Giordano, M.C. and G.F. acknowledge the PETASPIN association (www.petaspin.com). D.V. and E.R. also acknowledge funding from the project PON Capitale Umano (CIR_00030) funded by the Italian Ministry of Universities and Research (MUR), and D.D. n. 47 del 20 febbraio 2025 – “Assunzione di ricercatori internazionali post-dottorato”, CUP: D53C25000630007 (D.V.) and CUP: D53C25000620007 (E.R.).

**Data availability statement**

The data that support the findings of this study are available upon reasonable request from the corresponding author.

**Competing interests**

The authors declare no competing interests.

**Supporting Information**

Supporting Information is available from the author.